\documentclass[aps,pra,showpacs,twocolumn,superscriptaddress]{revtex4-1}
\usepackage{amsmath}
\usepackage{amssymb}
\usepackage{graphicx}
\usepackage{epstopdf}
\usepackage{times,txfonts}
\usepackage{easybmat}
\usepackage{mathtools}

\usepackage[bookmarks=false]{hyperref}
\hypersetup{colorlinks=true,citecolor=blue,linkcolor=blue,urlcolor=blue,pdfstartview=FitH,bookmarksopen=true}
\usepackage{color,soul}

\begin{document}
\title{A Hybrid Universal Blind Quantum Computation}
\author{Xiaoqian Zhang}
\affiliation{College of Information Science and Technology, Jinan University, Guangzhou 510632, China}
\author{Weiqi Luo}
\email{lwq@jnu.edu.cn}
\affiliation{College of Information Science and Technology, Jinan University, Guangzhou 510632, China}
\author{Guoqiang Zeng}
\email{zeng.guoqiang5@gmail.com}
\affiliation{College of Cyber Security, Jinan University, Guangzhou 510632, China}
\author{Jian Weng}
\affiliation{College of Information Science and Technology, Jinan University, Guangzhou 510632, China}
\author{\\ Yaxi Yang}
\affiliation{College of Information Science and Technology, Jinan University, Guangzhou 510632, China}
\author{Minrong Chen}
\affiliation{School of Computer, South China Normal University, Guangzhou 510631, China}
\author{Xiaoqing Tan}
\affiliation{Department of Mathematics, Jinan University, Guangzhou 510632, China}

\date{\today}

\begin{abstract}
  In blind quantum computation (BQC), a client delegates her quantum computation to a server with universal quantum computers who learns nothing about the client's private information. In measurement-based BQC model, entangled states are generally used to realize quantum computing. However, to generate a large-scale entangled state in experiment becomes a challenge issue. In circuit-based BQC model, single-qubit gates can be realized precisely, but entangled gates are probabilistically successful. This remains a challenge to realize entangled gates with a deterministic method in some systems. To solve above two problems, we propose the first hybrid universal BQC protocol based on measurements and circuits, where the client prepares single-qubit states and the server performs universal quantum computing. We analyze and prove the correctness, blindness and verifiability of the proposed protocol.
\end{abstract}

\maketitle

\section{Introduction}
Recently, blind quantum computation (BQC) becomes a hot topic in quantum information processing since it can be applied to realize clients' private quantum computing. In BQC, measurement-based model and circuit-based model have been studied for years
\cite{1Broadbent09,Barz12,2Morimae13,qian17,3Li14,4Sheng15,5Morimae2012,8Morimae2015,Yu18,Childs2005,22Fisher2014,Broa15,Mar16,Walther,xiao18}. A. Broadbent \emph{et al.} \cite{1Broadbent09} in 2009 firstly implemented a universal BQC protocol by measuring an $m\times n$ dimensional blind brickwork state, which is called Broadbent-Fitzsimons-Kashefi (BFK) protocol. In BFK protocol, the client can prepare single-qubit states $\{|\pm_\theta\rangle=\frac{1}{\sqrt{2}}(|0\rangle+e^{i\theta}|1\rangle)|\ \theta=0, \frac{\pi}{4}, \frac{2\pi}{4}, \ldots, \frac{7\pi}{4}\}$. Based on BFK protocol, multi-server BQC protocols were proposed in \cite{2Morimae13,3Li14,4Sheng15}. A BQC protocol for single-qubit gates X, Y, T, Z has been realized by measuring blind topological states, where the error threshold is explicitly calculated \cite{5Morimae2012}. A universal BQC protocol based on Affleck-Kennedy-Lieb-Tasaki (AKLT) states has been implemented, where the universal gates set consists of blind Z-rotation, blind X-rotation and controlled-Z followed by blind Z-rotations \cite{8Morimae2015}. In experiments, S. Barz \emph{et al.} \cite{Barz12} realized a demonstration for the privacy of quantum inputs, computations, and outputs. Furthermore, the verifiable BQC protocols and other interesting BQC protocols have been proposed \cite{7Morimae2014,6Hayashi2015,Gheor15,Fit15,Fu16,Moe16,Anne17,1Gio13,13Sun15,14Greg16,16Delg15,25Hua,30Huang}.
In \cite{13Sun15}, a blind quantum computing about symmetrically private retrieval was proposed, where a client Alice has limited quantum technologies and queries a item of the database owned by a server Bob who has a fledged quantum computer. In the protocol, the privacy of both participants can be preserved: Bob knows nothing about what Alice has retrieved, and Alice can only get the information that she wants to query of the database, where the related private retrieval schemes can refer to \cite{Gao15,Gao16,Wei18}.

For quantum computers, it is important to prepare entangled states that can be applied to quantum computing \cite{Rausse01}, quantum simulation \cite{Seth96} and so on. In measurement-based BQC model, the key problem is how to generate large-scale entangled states \cite{1Broadbent09,Horodecki09} in space-separated and individual-controllable quantum systems such as the brickwork state \cite{1Broadbent09}, AKLT state \cite{8Morimae2015}. In experiments, great progress has been made in preparing multi-qubit entangled states. The number of qubits in an entangled state \cite{Friis18} reaches to 20 in trapped-ion system, while the number is 10 both in superconducting \cite{Song17} and photonic systems \cite{Lin16}. It is difficult to describe a large-scale entangled state since the dimension of Hilbert space is exponentially increasing. In circuit-based BQC model \cite{Childs2005,22Fisher2014,Broa15,Mar16,Walther}, the entangled gates are realized probabilistically such as the successful probability in optical system is $1/16$ in \cite{Koashi01}, $1/9$ in \cite{Ralph02}, $1/4$ in \cite{Pittman01}, $1/3$ in \cite{Hofmann02} and $21/25$ in \cite{Brien03}.

In this paper, we first propose a hybrid universal BQC protocol (HUBQC), which is based on measurements and circuits. Intuitively, we make full use of advantages of two models. Specially, entangled gates can be realized with a deterministic method in measurement-based model, solving the probabilistic realization of entangled gates problem in circuit-based model. Meanwhile, the single-qubit gates can also be realized without too many qubits in circuit-based model, solving experimentally generation of a large-scale entangled state problem in measurement-based model. A client Alice generates initial states and a server Bob performs operations and measurements. The entangled gates can be realized by measuring graph states and single-qubit gates can be operated on the suitable qubits with an predefined order. We not only prove the correctness and blindness of the protocol but also have verifiability which implies to verify Bob's honesty and the correctness of measurement outcomes. Finally, we apply HUBQC protocol to realize blind quantum Fourier transform. For blindness, measurement process has adopted the encryption algorithm from BFK protocol.

The rest of this paper is organized as follows. We present the preliminaries in Section \ref{sec:jud}. The definition and structure of the graph state $|Cluster\rangle$ are presented in Section \ref{sec:stru}. The universal blind quantum computation protocol is in Section \ref{sec:pro}. We show the analyses and proofs of correctness, blindness and verifiability as well as a application of our protocol in Section \ref{sec:ana}. At last, our discussions and conclusions are given in Section \ref{sec:con}.

\section{Preliminaries}
\label{sec:jud}

\subsection{Basic principles of circuit-based quantum computation}

In \cite{2000MAN}, it points out that an arbitrary unitary operator U can be decomposed into the combinations of rotation operators. We first give the rotation operators as follows:
\begin{small}
\begin{eqnarray}
\begin{array}{l}
\displaystyle
R_x(\alpha)=\left(
  \begin{array}{cc}
  cos\frac{\alpha}{2}   &-isin\frac{\alpha}{2} \\
  -isin\frac{\alpha}{2} &  cos\frac{\alpha}{2}\\
  \end{array}
\right),\\
\displaystyle R_y(\beta)=\left(
  \begin{array}{cc}
 cos\frac{\beta}{2} &-sin\frac{\beta}{2} \\
  sin\frac{\beta}{2} &  cos\frac{\beta}{2}\\
  \end{array}
\right),\\
\displaystyle  R_z(\gamma)=\left(
  \begin{array}{cc}
  e^{\frac{-i\gamma}{2}} &0 \\
  0 & e^{\frac{i\gamma}{2}}\\
  \end{array}
\right),
\end{array}
\end{eqnarray}
\end{small}
where $\alpha, \beta, \gamma\in[0, 2\pi]$. Particularly, if the rotation angle is $\pi$ about $x$-axis, $y$-axis and $z$-axis respectively, we get
\begin{eqnarray}
R_x(\pi)=iX,\ R_y(\pi)=XZ,\ R_z(\pi)=-iZ.
\end{eqnarray}

If there exist $\theta$, $\alpha$, $\beta$ and $\gamma$, s.t. an arbitrary unitary operator U has the decompositions as follows:
\begin{small}
\begin{eqnarray}
\begin{array}{l}
\displaystyle U=e^{i\theta}R_z(\alpha)R_y(\beta)R_z(\gamma)\\
\displaystyle\qquad\qquad=\left(
  \begin{array}{cc}
  e^{i(\theta-\frac{\alpha}{2}-\frac{\gamma}{2})}cos\frac{\beta}{2} & -e^{i(\theta-\frac{\alpha}{2}+\frac{\gamma}{2})}sin\frac{\beta}{2} \\
  e^{i(\theta+\frac{\alpha}{2}-\frac{\gamma}{2})}sin\frac{\beta}{2} & e^{i(\theta+\frac{\alpha}{2}+\frac{\gamma}{2})}cos\frac{\beta}{2}\\
  \end{array}
\right),\\
\displaystyle U=e^{i\theta}R_z(\alpha)R_x(\beta)R_z(\gamma)\\
\displaystyle\qquad\qquad=\left(
  \begin{array}{cc}
  e^{i(\theta-\frac{\alpha}{2}-\frac{\gamma}{2})}cos\frac{\beta}{2} & -ie^{i(\theta-\frac{\alpha}{2}+\frac{\gamma}{2})}sin\frac{\beta}{2} \\
  -ie^{i(\theta+\frac{\alpha}{2}-\frac{\gamma}{2})}sin\frac{\beta}{2} & e^{i(\theta+\frac{\alpha}{2}+\frac{\gamma}{2})}cos\frac{\beta}{2}\\
  \end{array}
\right),\\
\displaystyle U=e^{i\theta}R_y(\alpha)R_x(\beta)R_y(\gamma)=e^{i\theta}\cdot \\
\displaystyle\left(
  \begin{array}{cc}
cos\frac{\beta}{2}cos\frac{\alpha+\gamma}{2}+isin\frac{\beta}{2}sin\frac{\alpha-\gamma}{2} & -cos\frac{\beta}{2}sin\frac{\alpha+\gamma}{2}-isin\frac{\beta}{2}cos\frac{\alpha-\gamma}{2} \\
cos\frac{\beta}{2}sin\frac{\alpha+\gamma}{2}-isin\frac{\beta}{2}cos\frac{\alpha-\gamma}{2} & cos\frac{\beta}{2}cos\frac{\alpha+\gamma}{2}-isin\frac{\beta}{2}sin\frac{\alpha-\gamma}{2}\\
  \end{array}
\right).
\end{array}
\end{eqnarray}
\end{small}

Here, we only show three decomposition forms, the other three decompositions $y\textnormal{-}z\textnormal{-}y$, $x\textnormal{-}z\textnormal{-}x$, $x\textnormal{-}y\textnormal{-}x$ are similar. Next, we give the $z\textnormal{-}y\textnormal{-}z$ decomposition for gates H, S, Z, T, X, Y as follows:
\begin{eqnarray}
\begin{array}{l}
\displaystyle H=e^{\frac{i \pi}{2}}R_y(\frac{\pi}{2})R_z(\pi),\ S=e^{\frac{i \pi}{4}}R_z(\frac{\pi}{2}),\ Z=e^{\frac{i \pi}{2}}R_z(\pi), \\
\displaystyle X=e^{\frac{i \pi}{2}}R_y(\pi)R_z(\pi),\ T=e^{\frac{i \pi}{8}} R_z(\frac{\pi}{4}), \ Y=e^{\frac{i\pi}{2}}R_y(\pi),
\end{array}
\end{eqnarray}

For the $z\textnormal{-}x\textnormal{-}z$ decomposition of rotation operators of above gates, we obtain
\begin{eqnarray}
\begin{array}{l}
\displaystyle  S=e^{\frac{i \pi}{4}}R_z(\frac{\pi}{2}),\quad Z=e^{\frac{i \pi}{2}}R_z(\pi),\quad T=e^{\frac{i \pi}{8}}R_z(\frac{\pi}{4}),\\
\displaystyle  X=e^{\frac{i \pi}{2}}R_x(\pi), \quad Y=e^{\frac{i\pi}{2}}R_x(\pi)R_z(\pi),\\
\displaystyle H=e^{\frac{i\pi}{2}}R_z(\frac{\pi}{2})R_x(\frac{\pi}{2})R_z(\frac{\pi}{2}).
\end{array}
\end{eqnarray}

For the $y\textnormal{-}x\textnormal{-}y$ decomposition of rotation operators of above gates, we get
\begin{eqnarray}
\begin{array}{l}
\displaystyle S=e^{\frac{i \pi}{4}}R_y(\frac{-\pi}{2})R_x(\frac{\pi}{2})R_y(\frac{\pi}{2}), \ H=e^{\frac{i\pi}{2}}R_x(\pi)R_y(\frac{\pi}{2}),\\
\displaystyle Z=e^{\frac{i \pi}{2}}R_y(\frac{-\pi}{2})R_x(\pi)R_y(\frac{\pi}{2}),\ \ X=e^{\frac{i\pi}{2}}R_x(\pi),\\
\displaystyle T=e^{\frac{i \pi}{8}}R_y(\frac{-\pi}{2})R_x(\frac{\pi}{4})R_y(\frac{\pi}{2}),\ Y=e^{\frac{i \pi}{2}}R_y(\pi).
\end{array}
\end{eqnarray}

Unexpected Pauli operators will appear in the process of circuit-based computation, therefore some main propagation relationships between rotation operators and Pauli operators can be expressed as follows:
\begin{eqnarray}
\begin{array}{l}
\displaystyle R_x(\beta)X=XR_x(\beta), \ \ \ R_x(\beta)Z=ZR_x(-\beta),\\
\displaystyle R_y(\beta)X=XR_y(-\beta), \ R_y(\beta)Z=ZR_y(-\beta),\\
\displaystyle R_z(\beta)X=XR_z(-\beta), \ R_z(\beta)Z=ZR_z(\beta).
\end{array}
\end{eqnarray}
Besides, the relationship of the rotation angles is $R_\phi(\alpha+\beta)=R_\phi(\alpha)\cdot R_\phi(\beta)$, where $\phi\in\{x, y, z\}$.

\subsection{Basic principles of measurement-based quantum computation}

In this section, we introduce the principles of measurement-based quantum computation.

In the paper \cite{1Broadbent09}, we first get the detailed definitions and technologies of single-qubit initial states, orthogonal projections measurements, gates corrections and two-qubit entanglement operators in measurement based quantum computation model. Second, if the measured qubits are not in the final column (vertical direction), the correction operations $X$, $Z$ and $R_z(\cdot)$ can be naturally absorbed by performing the adaptive projective measurements. Third, we also obtain the commutation relationships of Controlled-Z (CZ) with X, Z, $R_z(\cdot)$ in \cite{1Broadbent09}. The three points also can be found in \cite{Danos07,Jozsa05}. In addition, the commutation relationships of Pauli operators with $R_x(\cdot)$, $R_z(\cdot)$ are found in Eq.(7). After measuring the former qubit in a large graph state, the following gate will act on the latter qubit:
$W(\theta)=\frac{1}{\sqrt{2}}\left(
  \begin{array}{cc}
  1  & e^{i\theta} \\
  1  &  -e^{i\theta}\\
  \end{array}
  \right)=H\cdot P(\theta)$, where
$P(\theta)=\left(
  \begin{array}{cc}
  1  & 0 \\
  0  &  e^{i\theta}\\
  \end{array}
  \right).$

\section{The definition and structure of the graph state $|Cluster\rangle$}
\label{sec:stru}
\begin{figure}
  \centering
  \includegraphics[width=0.35\textwidth]{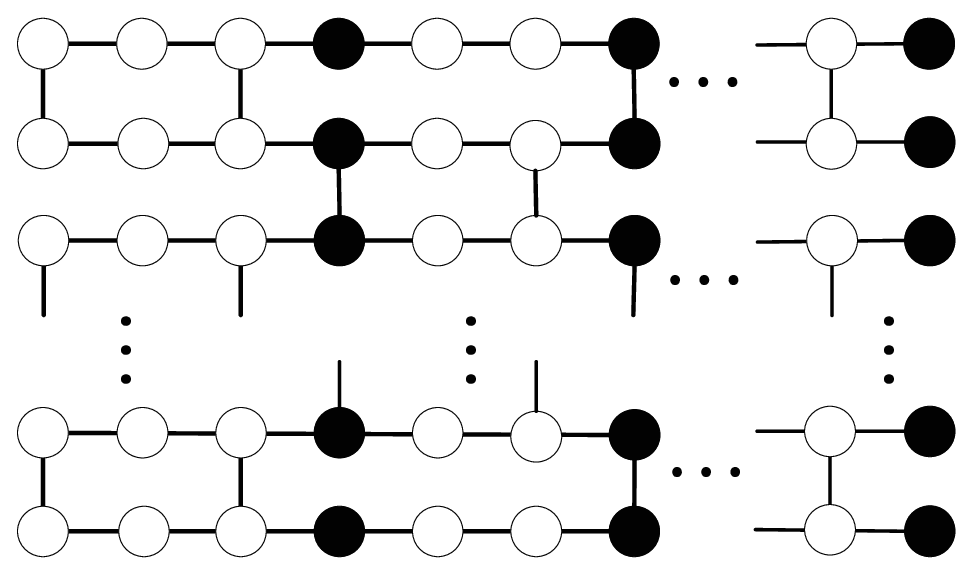}
  \caption{Schematic structure of a graph state $|Cluster\rangle$, where the black dots can be viewed as the outputs in former computing meanwhile the inputs in the latter computing. All white dots are auxiliary qubits to help complete the computing.}\label{A1}
\end{figure}

\emph{Definition}---In FIG. \ref{A1}, we show the structure of an $m \times n$ dimensional entangled state $|Cluster\rangle$, where these single-qubit states in the state $|Cluster\rangle$ are $|\pm_{\omega_j}\rangle=\frac{1}{\sqrt{2}}(|0\rangle\pm e^{i\omega_j}|1\rangle)$ ($\omega_j=0, \frac{\pi}{4}, \ldots, \frac{7\pi}{4}$). Suppose $m$ denote the horizontal rows and $n$ denote the vertical columns. The physical qubits are labelled as index $(a, b)$, where $a$ represents the $a$-th row and $b$ represents the $b$-th column.

1. For odd rows $a$ and columns $b\equiv 1$\ (mod\ 6), applying operations CZ on qubits $(a,b)$ and $(a+1,b)$, $(a,b+2)$ and $(a+1,b+2)$.

2. For even rows $a$ and columns $b\equiv 4$\ (mod\ 6), applying operations CZ on qubits $(a,b)$ and $(a+1,b)$, $(a,b+2)$ and $(a+1,b+2)$.

3. For each row $a$, applying operations CZ on qubits $(a, b)$ and $(a, b+1)$ where $1\leqslant a\leqslant m, 1\leqslant b\leqslant n$.

\begin{figure}
  \centering
  \includegraphics[width=0.5\textwidth]{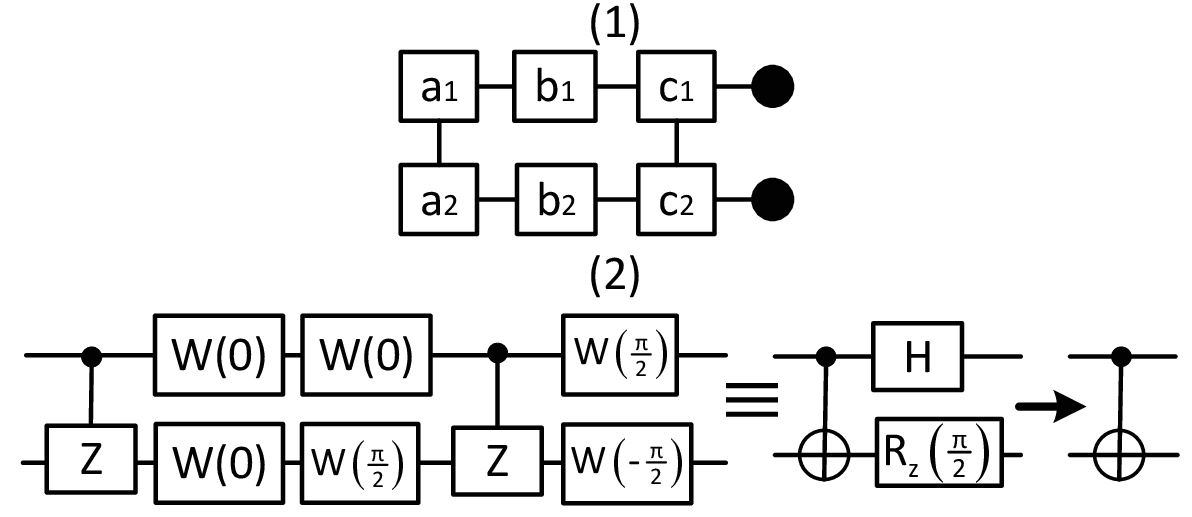}
  \caption{Schematic structure of an eight-qubit cluster state which refers to our previous work \cite{Xiao}, where qubits labelled by $a_f, b_f, c_f$ ($f=1, 2$) need to be measured. Except for a global phase factor, $W(\theta)$ is the same as $HR_z(\theta)$.}\label{A2}
\end{figure}

It can be seen from FIG. \ref{A1} that every unit state is an eight-qubit cluster state (See FIG. \ref{A2}(1)) which can be used to realize entangled gates Controlled-NOT (CNOT) (See FIG. \ref{A2}(2)).

\section{A hybrid universal BQC protocol}
\label{sec:pro}
\begin{figure}[!htp]
  \centering
  \includegraphics[width=0.4\textwidth]{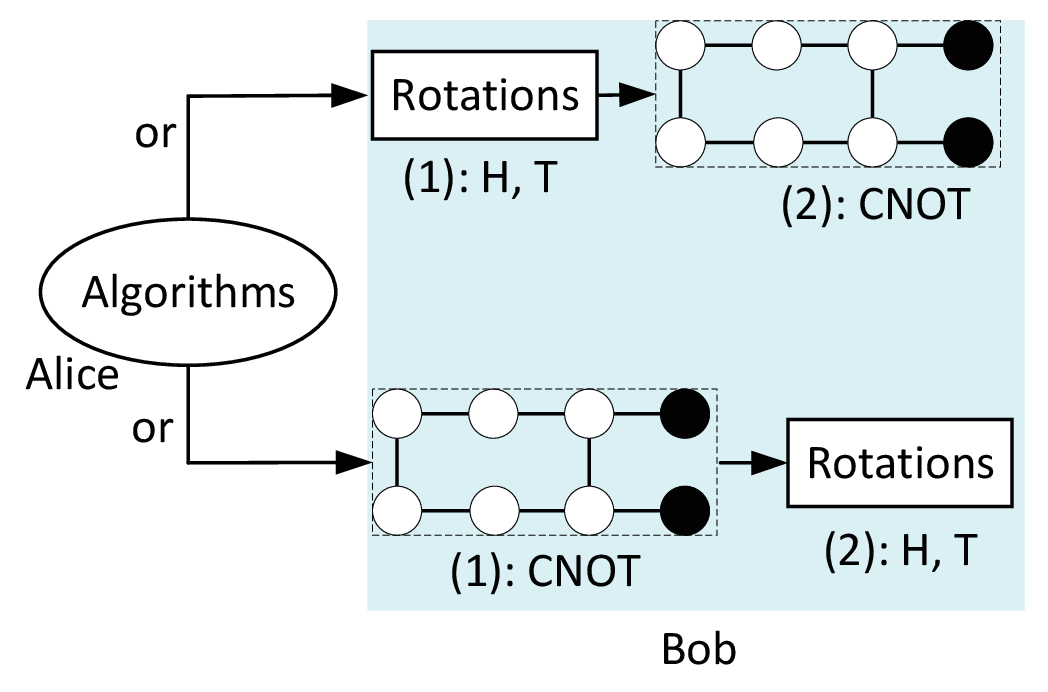}
  \caption{(Color online) Schematic diagram of our BQC protocol, where rotations denote the decompositions of gates H, T. The eight-qubit cluster state for realizing a CNOT gate belongs to state $|Cluster\rangle$. }\label{A3}
\end{figure}

\emph{Our HUBQC Protocol}---The concrete steps of our protocol are as follows (See FIG. \ref{A3}), where the client Alice has the ability to prepare the initial states and the server Bob can perform universal quantum computing without extracting Alice's any private information.

Step 1. Alice prepares all single-qubit states $|\pm_{\omega_j}\rangle$, $|0\rangle$, $|1\rangle$, $|\pm_{\mu_j}\rangle$ and sends them to Bob, where $\omega_j, \mu_j\in\{0, \frac{\pi}{4}, \cdots, \frac{7\pi}{4}\}$. These states $|\pm_{\omega_j}\rangle$ are used for computing and $|0\rangle$, $|1\rangle$, $|\pm_{\mu_j}\rangle$ are trap qubits. The reason choosing $|0\rangle$, $|1\rangle$, $|\pm_{\mu_j}\rangle$ as trap qubits is that $|0\rangle$, $|1\rangle$ are not entangled with $|\pm_{\mu_j}\rangle$ after performing CZ gates. While states $|\pm_{\mu_j}\rangle$ can be entangled with each other at most three qubits as long as they are in the suitable places. Note that, the connections with the states $|\pm_{\omega_j}\rangle$ are $|0\rangle$ and $|1\rangle$.

\begin{figure}
  \centering
  \includegraphics[width=0.35\textwidth]{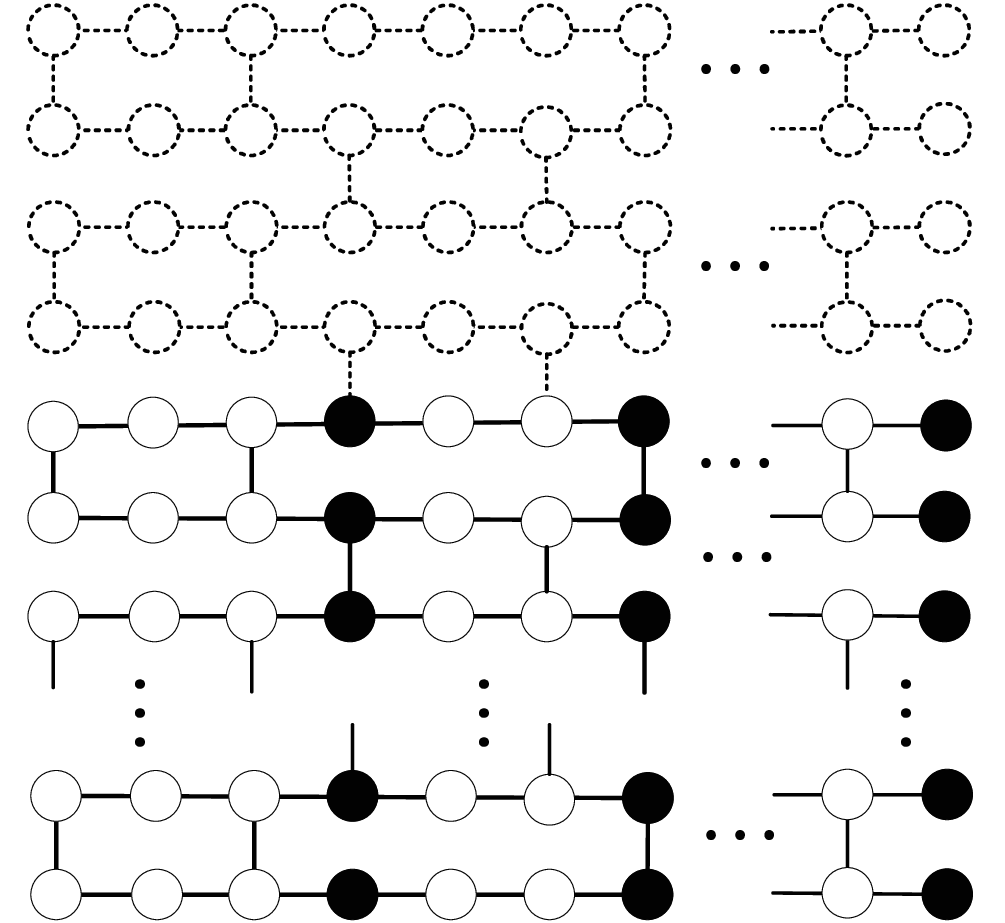}
  \caption{Schematic structure of an entangled state $|C\rangle$, where qubits connected by dotted lines are trap qubits $|0\rangle$, $|1\rangle$, $|\pm_{\mu_j}\rangle$ and solid lines are computational qubits $|\pm_{\omega_j}\rangle$. The positions of trap qubits are random without having an impact on the computing and keeping the structural consistency with computational qubits.}\label{A4}
\end{figure}

Step 2. Alice asks Bob to perform CZ gates to get eight-qubit cluster states and implement the corresponding measurements until Bob gets a graph state $|C\rangle$ (See Fig. \ref{A4}). In Fig. \ref{A4}, some qubits connected by dotted lines are trap qubits $|0\rangle$, $|1\rangle$, $|\pm_{\mu_j}\rangle$ and the others are computational qubits $|\pm_{_j}\rangle$. These trap qubits can be randomly attached to the $|Cluster\rangle$ state as long as they keep the structural consistency and do not affect the original computing.

Step 3. In Alice's target algorithms, if single-qubit gates are required to implement first, Alice asks Bob to perform the above process in FIG. \ref{A3}, where H and T are the combination of rotation operators. Bob first performs encrypted rotation operations on two black dots in the cluster state, where the encrypted rotation angles are $\xi_j=\nu_j+r_j\pi$ ($\nu_j$ is true rotation angles and $r_j$ is randomly chosen from $\{0, 1\}$) and $R_\phi(\xi_j)=R_\phi(r_j\pi+\nu_j)=R_\phi(r_j\pi)R_\phi(\nu_j)$ ($\phi\in\{x,y,z\}$). Note that, the encrypted angle $\xi_j$ and true rotation angles $\nu_j$ belong to the set $\{0, \frac{\pi}{4}, \frac{2\pi}{4}, \pi,\frac{5\pi}{4}, \frac{6\pi}{4}\}$.

Next, Bob measures every white dot qubit in the cluster state to get the CNOT gate, where the corresponding measurement angles are $\delta_t=\omega'_t+\kappa_t+r_t\pi$ which belongs to the set $\{0, \frac{\pi}{4}, \cdots, \frac{7\pi}{4}\}$. $r_t$ is randomly chosen from the set $\in\{0,1\}$, and $\omega'_t=(-1)^{s_t^X}\omega_t+s^Z_t\pi$ depends on previous measurement outcomes. The measurement results are zero in the first row and the first column \cite{1Broadbent09}.

Otherwise, Alice asks Bob to perform the below process in FIG. \ref{A3}. Bob first measures the white dots qubits to get a CNOT gate and then performs rotation operators in black dots qubits to realize a single-qubit gate. Note that for gates CNOT, if the cluster states do not contain final quantum outputs in FIG. \ref{A4}, the correction operations $R_z(\textnormal{-}\frac{\pi}{2})$ can be naturally absorbed by performing the projective measurements $|\pm_{\delta_t-\frac{\pi}{2}}\rangle$ since $|\pm_{\delta_t-\frac{\pi}{2}}\rangle$ is the same as $R_z(\textnormal{-}\frac{\pi}{2})|\pm_{\delta_t}\rangle=\frac{1}{\sqrt{2}}(e^{\frac{i\pi}{4}}|0\rangle\pm e^{i(\delta_t-\frac{\pi}{4})}|1\rangle)=\frac{e^{\frac{i\pi}{4}}}{\sqrt{2}}[|0\rangle\pm e^{i(\delta_t-\frac{\pi}{2})}|1\rangle]$ except for a global phase factor.

The above two processes can also be performed in trap qubits, therefore, Bob can not distinguish which are useful CNOT gates and trap gates CNOT in FIG. \ref{A4} to strengthened the security of our protocol.

Step 4. In the final quantum outputs, Alice asks Bob to perform the correct operations H and $R_z(\textnormal{-}\frac{\pi}{2})$. That is, Bob performs correct rotation operators $R_x(\cdot), R_y(\cdot)$ or $R_z(\cdot)$. After Bob returning all quantum outputs, Alice first measures the trap qubits to verify Bob's honesty, where the number of trap qubits is optimal without having an impact on the computational efficiency. In fact, eight-qubits cluster states can also be used to realize single-qubit gates \cite{Xiao}. Combined with trap gates and encoded measurement angles, it is impossible for Bob to know the position of CNOT gates.

In the protocol, Bob maybe implement Pauli attacks to change the original graph states. If Bob performs Pauli attacks X on $|0\rangle$, $|1\rangle$ or Z on $|\pm_{\mu_j}\rangle$ or XZ on $|0\rangle$, $|1\rangle$, $|\pm_{\mu_j}\rangle$, Alice will get violative results and she aborts the protocol. Note that, Alice knows all measurement results on traps with related basis. If Bob passes the verification, Alice will discard all traps and accept the results.

\section{Proofs and applications}
\label{sec:ana}
We first prove the correctness, blindness and verifiability of our HUBQC protocol.

\emph{Correctness.} All quantum outputs are correct when Bob performs the protocol honestly.

\begin{figure}
  \centering
  \includegraphics[width=0.4\textwidth]{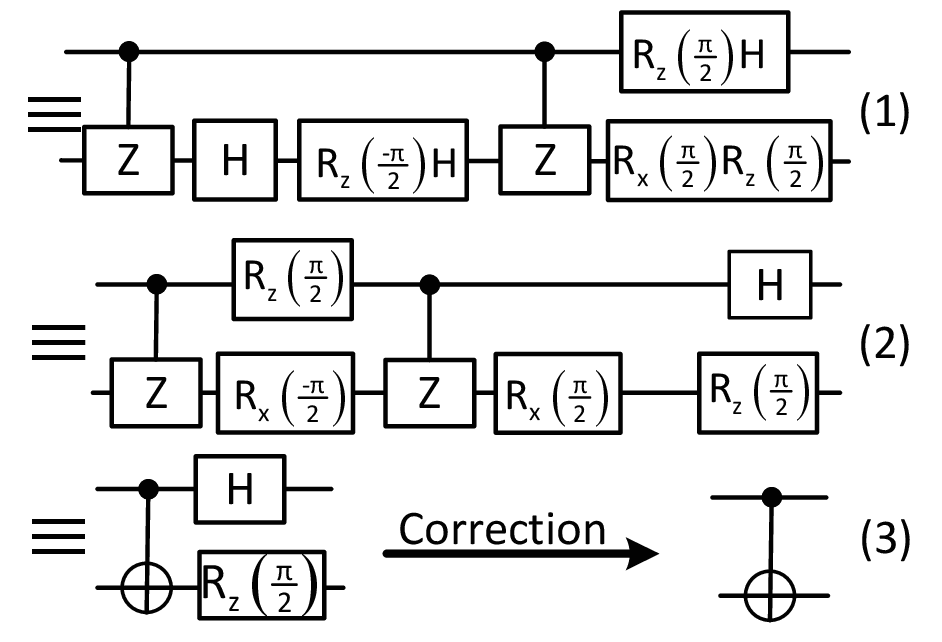}
  \caption{The simplification process of CNOT gate.}\label{A5}
\end{figure}

\emph{Proof:}
1) In measurement-based process, the correctness of gate CNOT is showed in FIG. \ref{A5}.

Since $H=e^{\frac{i\pi}{2}}R_z(\frac{\pi}{2})R_x(\frac{\pi}{2})R_z(\frac{\pi}{2})$ holds, we get $R_z(\textnormal{-}\frac{\pi}{2})H=e^{ \frac{i\pi}{2}}R_x(\frac{\pi}{2})R_z(\frac{\pi}{2})$ in the below lines. After that, we obtain the circuit (1). And we get the circuit (2) via the relationship $HR_z(\alpha)H=R_x(\alpha)$. By correcting H and $R_z(\textnormal{-}\frac{\pi}{2})$, we receive the gate CNOT with the relationship $(R_z(\frac{\pi}{2})\otimes R_x(\frac{\pi}{2}))CZ(I\otimes R_x(\textnormal{-}\frac{\pi}{2}))CZ=CNOT$.

In the circuit process, the correctness can also be ensured since we have
\begin{eqnarray*}
\begin{array}{l}
\displaystyle
R_x(\nu_j+r\pi)=
\begin{cases}
R_x(\nu_j), &r=0\cr iXR_x(\nu_j), &r=1
\end{cases},\\
\displaystyle R_y(\nu_j+r\pi)=
\begin{cases}
R_y(\nu_j), &r=0\cr XZR_y(\nu_j), &r=1
\end{cases},\\
\displaystyle R_z(\nu_j+r\pi)=
\begin{cases}
R_z(\nu_j), &r=0\cr -iZR_z(\nu_j), &r=1
\end{cases},
\end{array}
\end{eqnarray*}
where X, Z are commuted with rotation operations so they can be easily removed. $\square$

\emph{Blindness (quantum inputs).} Suppose the quantum inputs are single-qubit states $|\pm_{\theta_j}\rangle$, $|0\rangle$, $|1\rangle$. Bob can not get anything from these qubits since the density matrices are maximally mixed from his point of view.

\emph{Proof:} For single-qubit states $|\pm_{\theta_j}\rangle$ and $|0\rangle$, $|1\rangle$, where $\theta_j\in\{0, \frac{\pi}{4}, \cdots, \frac{7\pi}{4}\}$, we have
\begin{eqnarray}
\begin{array}{l}
\displaystyle \frac{1}{18}[\sum\nolimits_{\theta_j}[|+_{\theta_j}\rangle\langle+_{\theta_j}|+|-_{\theta_j}\rangle\langle-_{\theta_j}|+|0\rangle\langle0|+|1\rangle\langle1|]\\
\displaystyle =\frac{1}{18}[|+\rangle\langle+|+|+_{\frac{\pi}{4}}\rangle\langle+_{\frac{\pi}{4}}|+\cdots+|+_{\frac{7\pi}{4}}\rangle\langle+_{\frac{7\pi}{4}}|\\
\displaystyle \qquad\ +|-\rangle\langle-|+|-_{\frac{\pi}{4}}\rangle\langle-_{\frac{\pi}{4}}|+\cdots+|-_{\frac{7\pi}{4}}\rangle\langle-_{\frac{7\pi}{4}}|\\
\displaystyle \qquad\ +|0\rangle\langle0|+|1\rangle\langle1|]=\frac{1}{2}I.
\end{array}
\end{eqnarray}
From the equation, we can get the conclusion: the density matrix is independent of quantum inputs, that is, Bob get nothing from the initial states.$\square$

\emph{Blindness (graph states).} The graph state $|C\rangle$ is completely blind including the dimension since it contains trap qubits.

\emph{Proof:}
Suppose the dimension of the graph state $|C\rangle$ is $m\times n$ known by Bob. However, the true dimension of state $|Cluster\rangle$ is smaller than $m\times n$. All units are eight-qubit cluster states, so nothing about the structure of state $|C\rangle$ is leaked. And the number and the positions of CNOT gates are secret for Bob. Moreover, all measurement angles are encrypted by one-time-pad. Therefore, Bob knows nothing about Alice's quantum computing.$\square$

\emph{Blindness (algorithms and outputs).} Here, two cases are considered: measurement-based process and circuit-based process. Bayes' theorem can be used to prove the blindness of quantum algorithms and outputs: $\bf{a)}$ the conditional probability distribution of computational angles known by Bob is equal to its priori probability distribution, when Bob knows some classical information and measurement outcomes of any positive-operator valued measurements (POVMs) at any time; $\bf{b)}$ all quantum outputs are one-time padded to Bob.

\emph{Proof:} In measurement-based process, the encrypted form is the same as the BFK protocol \cite{1Broadbent09}, the blindness proofs of algorithms and outputs are also the same as those in \cite{5Morimae2012,8Morimae2015}. In circuit-based process, the encrypted form is $\xi_j=\nu_j+r\pi$, we give the blindness proofs of algorithms and outputs as follows.

We firstly analyse the effect of Bob's rotation angles information $\Xi_j=\{\xi_j\}_{j=1}^m$ on Alice's privacy \cite{5Morimae2012,8Morimae2015}. Suppose $V_j=\{\nu_j\}_{j=1}^m$, $R_j=\{r_j\}_{j=1}^m$, where $R_j\in\{0, 1\}$ is a random variable chosen by Alice and $\{\Xi_j, V_j \}\in S=\{\frac{k\pi}{4}\mid k=0, 1, 2, 4, 5, 6 \}$. Let $\Lambda\in\{1,\cdots,m\}$ be a random variable related with an operation.
The conditional probability distribution of $\Xi_j$ given by $\Lambda=j$ and $V_j$ shows Bob's knowledge which is about Alice's rotation angles information. Based on Bayes' theorem, we get
\begin{widetext}
\begin{small}
\begin{eqnarray*}
\begin{array}{l}
\displaystyle p(\Xi_j=\{\xi_j\}_{j=1}^m \mid \Lambda=j,V_j=\{\nu_j\}_{j=1}^m)\\
\displaystyle =\frac{p(\Lambda=j\mid \Xi_j=\{\xi_j\}_{j=1}^m,V_j=\{\nu_j\}_{j=1}^m)p(\Xi_j=\{\xi_j\}_{j=1}^m,V_j=\{\theta_j\}_{j=1}^m)}{p(\Lambda=j,V_j=\{\nu_j\}_{j=1}^m)}\\
\displaystyle =\frac{p(\Lambda=j\mid \Xi_j=\{\xi_j\}_{j=1}^m,V_j=\{\nu_j\}_{j=1}^m)p(\Xi_j=\{\xi_j\}_{j=1}^m)p(V_j=\{\nu_j\}_{j=1}^m )}{p(\Lambda=j\mid V_j=\{\nu_j\}_{j=1}^m)p(V_j=\{\nu_j\}_{j=1}^m)}\\
\displaystyle =p(\Xi_j=\{\xi_j\}_{j=1}^m)\cdot
\frac{p(\Lambda=j\mid \Xi_j=\{\xi_j\}_{j=1}^m,V_j=\{\nu_j\}_{j=1}^m)}{p(\Lambda=j\mid V_j=\{\nu_j\}_{j=1}^m)}\\
\displaystyle =p(\Xi_j=\{\xi_j\}_{j=1}^m).
\end{array}
\end{eqnarray*}
\end{small}
\end{widetext}
\noindent This implies that the conditional probability distribution of rotation angles known by Bob is equal to its priori probability distribution. So our HUBQC protocol satisfies the condition $\bf{a)}$.

Similarly, we can get the conditional probability as follows:
\begin{widetext}
\begin{small}
\begin{eqnarray*}
\begin{array}{l}
\displaystyle p(R_j=\{r_j\}_{j=1}^m \mid \Lambda=j,\Xi_j=\{\xi_j\}_{j=1}^m)\\
\displaystyle=\frac{p(\Lambda=j\mid R_j=\{r_j\}_{j=1}^m,V_j=\{\nu_j\}_{j=1}^m)p(R_j=\{r_j\}_{j=1}^m,\Xi_j=\{\xi_j\}_{j=1}^m)}{p(\Lambda=j,\Xi_j=\{\xi_j\}_{j=1}^m)}\\
\displaystyle =\frac{p(\Lambda=j\mid R_j=\{r_j\}_{j=1}^m,\Xi_j=\{\xi_j\}_{j=1}^m)p(R_j=\{r_j\}_{j=1}^m)p(\Xi_j=\{\xi_j\}_{j=1}^m)}{p(\Lambda=j\mid V_j=\{\nu_j\}_{j=1}^m)p(\Xi_j=\{\xi_j\}_{j=1}^m)}\\
\displaystyle =p(R_j=\{r_j\}_{j=1}^m)\frac{p(\Lambda=j\mid R_j=\{r_j\}_{j=1}^m,\Xi_j=\{\xi_j\}_{j=1}^m)}{p(\Lambda=j\mid \Xi_j=\{\xi_j\}_{j=1}^m)}\\
\displaystyle =p(R_j=\{r_j\}_{j=1}^m).
\end{array}
\end{eqnarray*}
\end{small}
\end{widetext}

The result shows that the value $\{r_j\}_{j=1}^m$ is independent of $\Xi_j=\{\xi_j\}_{j=1}^m$, so our HUBQC protocol satisfies the condition $\bf{b)}$.$\square$

\emph{Verifiability.} The verifiability is to ensure that the client Alice can obtain the correct results and the server Bob is honest. That is, if all measurements on traps show the correct results, the probability that a logical state of Alice's computation is changed is exponentially small.

\emph{Proof:}
In our protocol, Alice adds some trap qubits around the state $|Cluster\rangle$. Bob knows neither the number of trap qubits nor their positions. When Bob returns these results, Alice makes a comparison between true results and Bob's results on the trap qubits. If the error rate is acceptable, Alice accepts these results on computational qubits. Moreover, Alice can measure the quantum outputs traps, and then successfully verifies Bob's honesty and the correctness of quantum computing.

Bob replaces the true $|C\rangle$ state with any states $\rho$. This equals to that Bob performs Pauli attacks I, X, Z, XZ. The proof is as follows, which refers to \cite{7Morimae2014}.

Now we show that the probability that Alice is fooled by Bob is exponentially small. Since Bob might be dishonest, he will deviate from the correct steps. His general attack is a creation of a different state $\rho$ instead of $|C\rangle$. If he is honest, $\rho=|C\rangle\langle C|$. If he is not honest, $\rho$ can be any state. The case can be deduced to Pauli attacks by a completely positive-trace-preserving (CPTP) map, and the details can refer to \cite{7Morimae2014}.

Suppose the qubits number of state $|C\rangle$ is $2N$, where the number of traps and computational qubits is $N$ respectively. Here, we denote that the number $N$ is optimal for traps. Then, the probability that all X operators of $\sigma_\alpha$ do not change any trap is $\frac{(2N-a)!\Pi_{k=0}^{a-1}(\frac{N}{2}-k)}{(2N)!}=(\frac{1}{2})^a\frac{\Pi_{k=0}^{a-1}(N-2k)}{\Pi_{k=0}^{a-1}(2N-k)}\leqslant (\frac{1}{2})^a\leqslant (\frac{1}{2})^{\alpha/3}$. We can obtain the same result for $max(a,b,c)=b$. For $max(a,b,c)=c$, we have $\frac{(2N-a)!\Pi_{k=0}^{a-1}(N-k)}{(2N)!}=\frac{\Pi_{k=0}^{a-1}(N-k)}{\Pi_{k=0}^{a-1}(2N-k)}\leqslant (\frac{1}{2})^a\leqslant (\frac{1}{2})^{\alpha/3}$. It implies that the probability that Alice is fooled by Bob is exponentially small. Hence our protocol is verifiable.$\square$

\begin{figure}[!htp]
  \centering
  \includegraphics[width=0.5\textwidth]{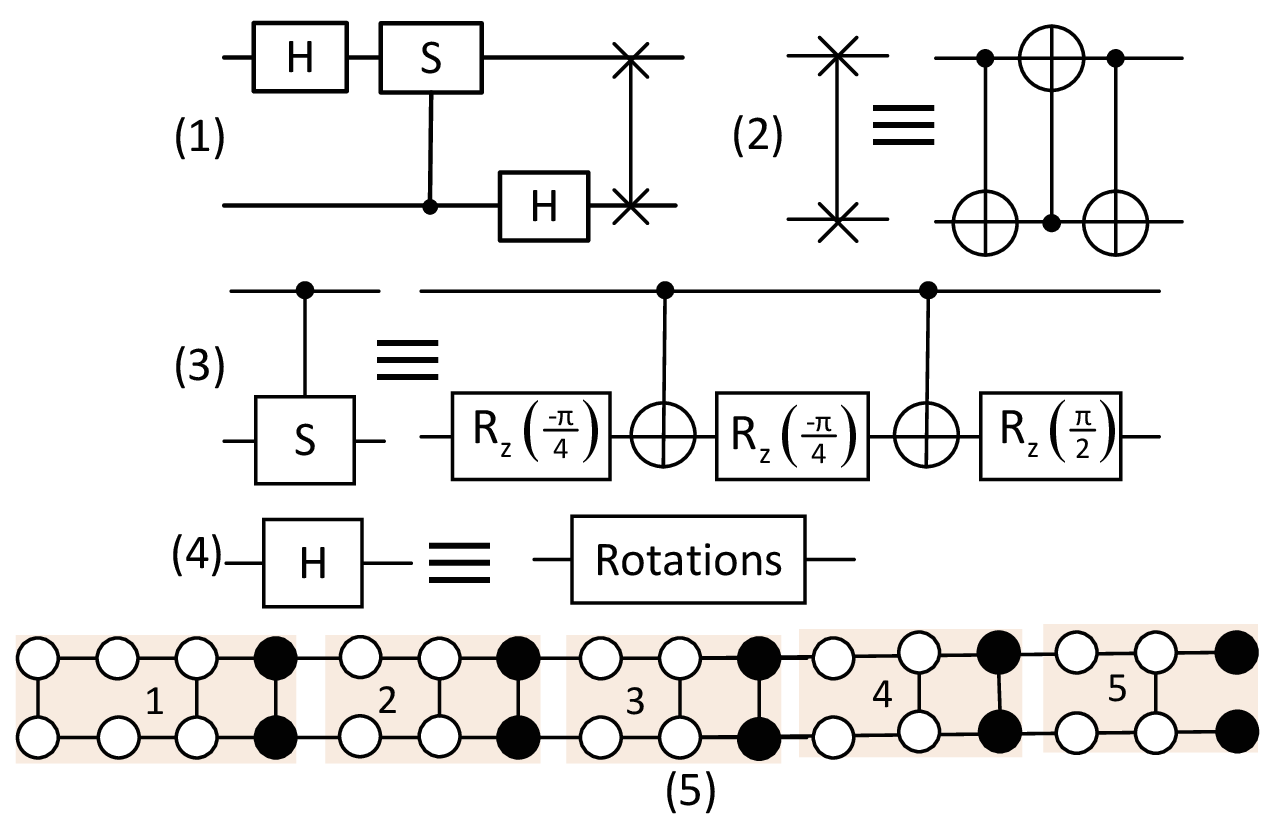}
  \caption{(Color online) (1) The quantum circuit is two-qubit QFT, where (2) shows the decomposition of SWAP gate, and (3) exhibits the decomposition of controlled-S, and (4) gives the combination of gate H, and (5) shows the structure of the graph state for realizing five CNOT gates, where trap qubits are not considered.}\label{A6}
\end{figure}

\emph{Application (Blind quantum Fourier transform)}---With the help of our HUBQC protocol, we study the quantum Fourier transform (QFT)  \cite{Marq10,Nam14,Lid17} and show the corresponding blind QFT protocol since multi-qubit QFT are the combinations of some single-qubit gates and entangled gates orderly.

We first explain how to realize blind two-qubit QFT. In FIG. \ref{A6}, all gates can be decomposed into rotation operations and CNOT gates. In \cite{2000MAN}, the decomposition principle of every controlled unitary operator $U$ has been given. For the unitary operator $U$, there are unitary operators A, B, C such that $ABC=I$ and $U=e^{i\alpha}AXBXC$, where $\alpha$ is a global phase factor. Suppose $A=R_z(\beta)R_y(\frac{\gamma}{2}), B=R_y(\textnormal{-}\frac{\gamma}{2})R_z(\frac{\textnormal{-}(\delta+\beta)}{2}), C=R_z(\frac{(\delta-\beta)}{2})$, $U=S$, we have
$\small
S=e^{i\alpha}R_z(\beta)R_y(\gamma)R_z(\delta)=\left(
  \begin{array}{cc}
  1   &  0 \\
  0   &  i\\
  \end{array}
\right)$. Set $\alpha=\frac{\pi}{4}$, $\beta=\frac{\pi}{2}$ and $\gamma=\delta=0$, so we get the Fig. \ref{A6}(3) about the decomposition of controlled-S entangled gate.

We also give the multi-qubit QFT referred to \cite{2000MAN} and the corresponding blind QFT protocol also can be realized via a similar way, where gate controlled-$G_n$ can also be decomposed into a combination of rotation operations and CNOT gates. Let $U=G_n$, we have
$\small
U=\left(
  \begin{array}{cc}
  1   &  0 \\
  0   &  e^{\frac{2\pi i}{2^k}}\\
  \end{array}
\right)$.
We set $\alpha=\frac{\pi}{2^k}$, $\gamma=0$ and $\beta+\delta=\frac{2\pi}{2^k}$.

\section{Discussions and Conclusions}
\label{sec:con}
In this section, we will discuss the measurement-based universal BQC , circuit-based universal BQC and our proposed HUBQC protocols.

$\bullet$ In measurement-based universal BQC model \cite{1Broadbent09}, every gate needs ten-qubit cluster states. So it brings a challenge to generate multi-qubits entangled states in experiments. In our protocol, we can divide the universal BQC protocol into two processes: measurement-based process and circuit-based process. We do not need a large-scale entangled state since only entangled gate need to be realized by using cluster states.

$\bullet$ In circuit-based universal BQC model \cite{Koashi01,Ralph02,Pittman01,Hofmann02,Brien03}, entangled gates in some systems are probabilistically successful, while the cluster states can be to determinately realize entangled gates.

$\bullet$ In our HUBQC protocol, compared with other works \cite{7Morimae2014,Moe16}, Alice has less workload since she only needs to measure trap qubits appearing in the final column of the graph state (See Fig. \ref{A4}). In measurement-based process, $\omega'_t+\kappa_t$ represents an actual measurement angle and $r_t$ is randomly chosen from $\{0, 1\}$ in $\delta_t=\omega'_t+\kappa_t+r_t\pi$. However, in circuit-based process, $r_j$ is also randomly chosen from $\{0, 1\}$ such that $\xi_j$ can be mapped to a uniform distribution set. In both processes, quantum outputs are all encrypted.

In summary, we propose a universal blind quantum computation protocol based on measurements and circuits which only needs two participants: a client Alice and a server Bob. Alice prepares the initial states and sends to Bob who creates the entangled state. According to the computations, Alice asks Bob to perform single-qubit rotation operators or entangled gates. Since the graph state $|Cluster\rangle$ is surrounded by many traps, and the structure of traps is the same as that of computational qubits, the state $|C\rangle$ is blind from Bob's perspective. In both measurement-based process and the circuit-based process, we encrypt the measurement angles and the rotation angles by one-time-pad. The correctness, blindness and verifiability have already been proved and the universality is obvious since the gates set is H, T, CNOT in our protocol.

\section*{Acknowledgments}
This work was supported by National Key R\&D Plan of China (Grant No. 2017YFB0802203, 2018YFB1003701), National Natural Science Foundation of China (Grant Nos. 61825203, 61872153, 61877029, 61872153, 61802145, U1736203, 61472165, 61732021, U1636209, 61672014), National Joint Engineering Research Center of Network Security Detection and Protection Technology, Guangdong Provincial Special Funds for Applied Technology Research and Development and Transformation of Important Scientific and Technological Achieve (Grant Nos. 2016B010124009 and 2017B010124002), Natural Science Foundation of Guangdong Province (2018A030313318), Guangdong Key Laboratory of Data Security and Privacy Preserving (Grant No. 2017B030301004), Guangzhou Key Laboratory of Data Security and Privacy Preserving (Grant No. 201705030004), National Cryptography Development Fund MMJJ20180109, and the Fundamental Research Funds for the Central Universities.

\nocite{*}
\bibliographystyle{elsarticle-num}
\bibliography{apssamp}

\end{document}